\documentclass[journal]{IEEEtran}
\usepackage{cite,amsmath}
\ifCLASSOPTIONcompsoc
\usepackage[caption=false, font=normalsize, labelfont=sf, textfont=sf]{subfig}
\else
\usepackage[caption=false, font=footnotesize]{subfig}
\fi
\usepackage{amsmath}
\usepackage{color}
\usepackage{graphicx}
\usepackage{pstool}
\usepackage{makecell}
\usepackage{amsfonts} 
\usepackage{dblfloatfix} 

\newcommand{\ve}[1]{\mathbf{#1}}
\newcommand{\te}[1]{\overline{\overline{#1}}}

\usepackage{threeparttable}
\usepackage{amsfonts}
\usepackage{amssymb}
\usepackage{multirow}
\usepackage{float}

\usepackage{graphicx,bm}
\usepackage[usenames,dvipsnames]{xcolor}
\usepackage{subfig}
\usepackage{cancel}
\usepackage{pstool,pstricks}

\begin{document}

\title{Multipolar Origin of the Unexpected Transverse Force Resulting from Two-Wave Interference}

\author{Karim~Achouri, Andrei Kiselev, and Olivier J. F. Martin}

\maketitle

\begin{abstract}
We propose a theoretical study on the electromagnetic forces resulting from the superposition of a TE and TM plane waves interacting with a sphere. Specifically, we first show that, under such an illumination condition, the sphere is subjected to a force transverse to the propagation direction of the waves. We then analyze the physical origin of this counter-intuitive behavior using a multipolar decomposition of the electromagnetic modes involved in that scattering process. This analysis reveals that interference effects, due to the two-wave illumination, lead to a Kerker-like asymmetric scattering behavior resulting in this peculiar transverse force.
\end{abstract}

\maketitle

\section{Introduction}

Electromagnetic forces, especially in the optical regime, have been the source of many studies and have accordingly generated tremendous attention~\cite{jonasLightWorkUse2008,zemanekPerspectiveLightinducedTransport2019}. Remarkably, they have been leveraged for various concepts and applications both at the nanoscopic scale for nanoparticle trapping, moving and sorting~\cite{ashkinAccelerationTrappingParticles1970,ashkin1986observation,neumanOpticalTrapping2004,wangMicrofluidicSortingMammalian2005,yangOpticalManipulationNanoparticles2009,juan2011plasmon,dholakia2011shaping,pang2012optical}, and at the macroscopic scale for solar sail steering~\cite{Johnson2011,achouriSolarMetaSailsAgile2019,swartzlanderLightSailingGreat2020}.

\begin{figure}[h]
	\centering
	\includegraphics[width=0.8\linewidth]{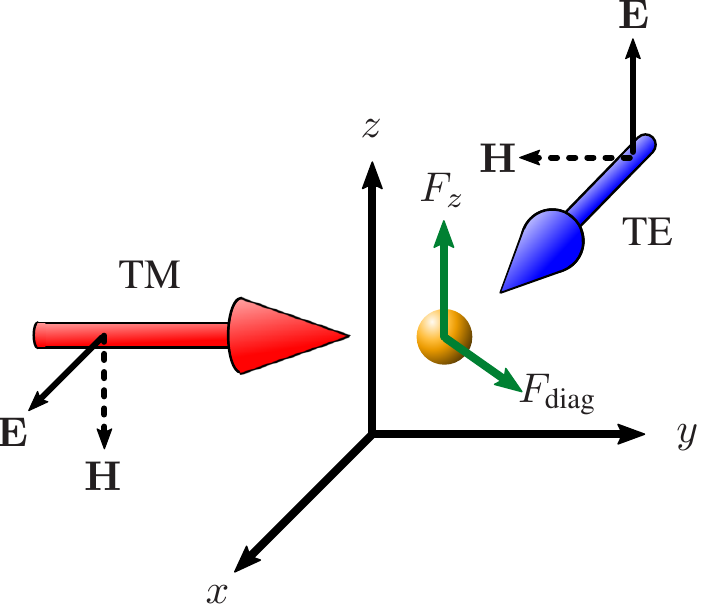}
	\caption{Forces acting on a metallic sphere when illuminated by a superposition of TE and TM waves. Due to this particular illumination condition, an unexpected transverse force $F_z$ appears in the direction normal to the illumination plane.}
	\label{fig:Schem}
\end{figure}

From an intuitive perspective, the origin of these forces may generally be explained by invoking the conservation of momentum or the presence of a field intensity gradient~\cite{radescuExactCalculationAngular2002, rothwell2008electromagnetics, salandrinoGeneralizedMieTheory2012,Novotny2012, chenOpticalPullingForce2011}. However, there are situations where the optical force results from much less intuitive and known phenomena. This is for instance the case when more than one illumination is considered, such as the two-wave interference cases discussed in~\cite{bliokhExtraordinaryMomentumSpin2014,bekshaevTransverseSpinMomentum2015,bliokhTransverseLongitudinalAngular2015,antognozziDirectMeasurementsExtraordinary2016,mobiniTheoryOpticalForces2018,gurvitzNumericalCalculationCartesian2018}, and which is the main topic of this work. In these cases, the combined effects of a two-wave illumination scheme results in peculiar outcomes such as the presence of an electromagnetic force transverse to the direction of wave propagation. This exotic effect is illustrated in Fig.~\ref{fig:Schem}, where a metallic sphere is illuminated by the superposition of a TE and a TM polarized waves.

In this configuration, the sphere is subjected to an expected in-plane diagonal force $F_\text{diag}$, as if the two waves were ``pushing'' on the sphere, but also to a surprising transverse force $F_z$, whose origin remains bewildering.

The purpose of this work is to shine some light on the physical origin of this transverse force and provide the reader with an intuitive and visual explanation based on a multipolar analysis using similar methods as in~\cite{nieto2010optical,chenOpticalPullingForce2011,evlyukhinCollectiveResonancesMetal2012}. This contrasts with other theoretical works on this topic, such as those in~\cite{bliokhExtraordinaryMomentumSpin2014,bekshaevTransverseSpinMomentum2015,bliokhTransverseLongitudinalAngular2015,antognozziDirectMeasurementsExtraordinary2016,mobiniTheoryOpticalForces2018,gurvitzNumericalCalculationCartesian2018}, where similar transverse forces have been reported but addressed in more abstract fashion. In order to remain succinct and thus avoid complicated considerations, we will concentrate our attention on the case of obliquely propagating TE and TM plane waves, as in Fig.~\ref{fig:Schem}, and study the origin of the transverse optical force acting on a metallic sphere.

\begin{figure*}[!b]
	\centering
	\subfloat[]{\label{fig:SIM1}
		\includegraphics[width=0.4\linewidth]{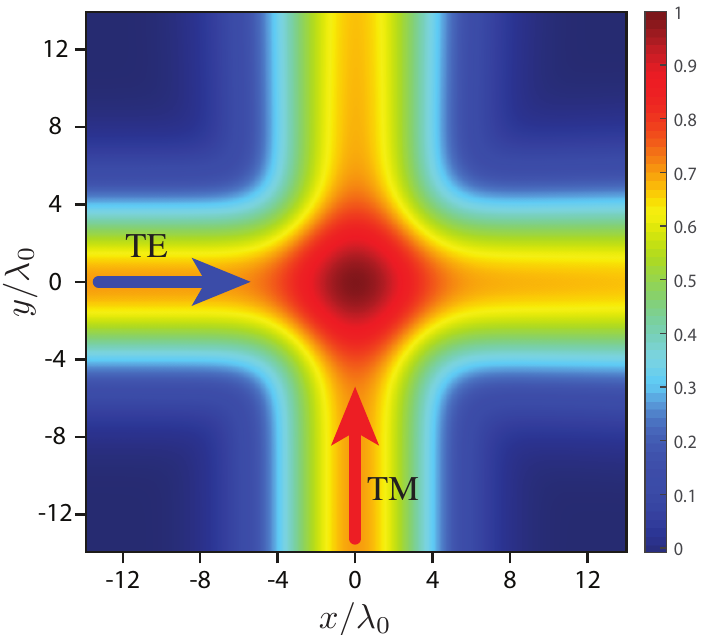}\label{fig:Abs}
	}\qquad\qquad\qquad
	\subfloat[]{\label{fig:SIM2}
		\includegraphics[width=0.4\linewidth]{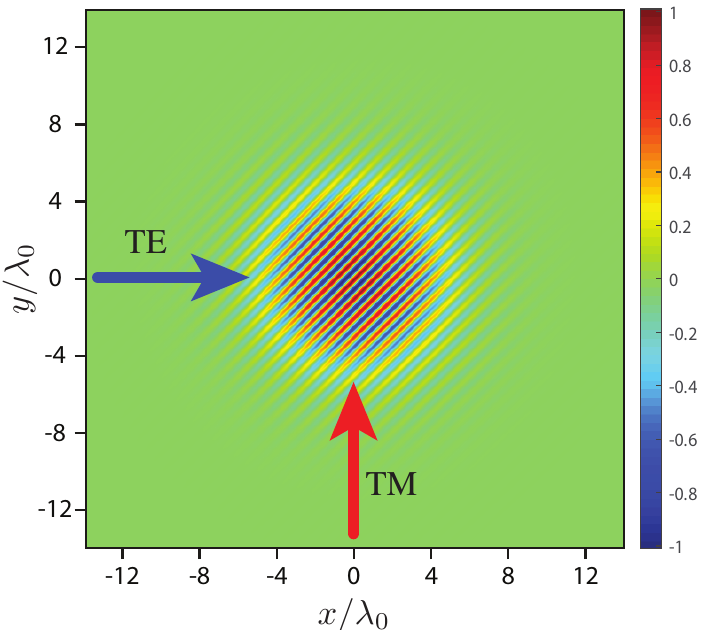}\label{fig:Sy}
	}
	\caption{Superposition of TE and TM Gaussian beams propagating in the $xy$-plane, as in Fig.~\ref{fig:Schem}. (a) Normalized electric field intensity. (b) Normalized time-averaged $z$-component of the Poynting vector. The arrows indicate the propagation direction of the two waves.}
	\label{fig:SIM}
\end{figure*}

\section{Superposition of Oblique TE and TM Waves}

Let us consider the propagation of a TE and a TM plane waves in the $xy$-plane. Note that the TE and TM polarizations are here defined with respect to the $xy$-plane. They propagate obliquely at angles $\phi_\text{TE}$ and $\phi_\text{TM}$ measured from the $x$-axis, respectively. The corresponding electric and magnetic field components, assuming propagation in vacuum and time dependence $e^{j\omega t}$, are given by
\begin{subequations}
	\label{eq:EFields}
	\begin{align}
	E_x &= -A_\text{TM} \frac{k_{\text{TM},y}}{k_0} e^{-j\ve{k}_\text{TM}\cdot\ve{r}},\\
	E_y &= +A_\text{TM} \frac{k_{\text{TM},x}}{k_0} e^{-j\ve{k}_\text{TM}\cdot\ve{r}},\\
	E_z &= +A_\text{TE}  e^{-j\ve{k}_\text{TE}\cdot\ve{r}+j\alpha},
	\end{align}
\end{subequations}
and
\begin{subequations}
	\label{eq:HFields}
	\begin{align}
	H_x &= +\frac{A_\text{TE}}{\eta_0} \frac{k_{\text{TE},y}}{k_0} e^{-j\ve{k}_\text{TM}\cdot\ve{r} + j\alpha},\\
	H_y &= -\frac{A_\text{TE}}{\eta_0} \frac{k_{\text{TE},x}}{k_0} e^{-j\ve{k}_\text{TM}\cdot\ve{r} + j\alpha},\\
	H_z &= +\frac{A_\text{TM}}{\eta_0} e^{-j\ve{k}_\text{TM}\cdot\ve{r}},
	\end{align}
\end{subequations}
where $A_\text{TE}$ and $A_\text{TM}$ are respectively the real amplitude of the TE and TM waves, $\eta_{0}$ and $k_{0}$ are the impedance and wavenumber in free space, $\alpha$ is the phase-shift between the two waves, $\ve{k}_\text{TE/TM} = k_0\left[\cos{(\phi_\text{TE/TM})}\ve{\hat{x}}+\sin{(\phi_\text{TE/TM})}\ve{\hat{y}}\right]$ and $\ve{r} = x\ve{\hat{x}}+y\ve{\hat{y}}$.

An interesting consequence of the superposition of these two plane waves becomes apparent when computing the corresponding total time-averaged Poynting vector, which reads
\begin{equation}
\label{eq:Poynting}
\begin{split}
&\langle\ve{S}\rangle=\frac{1}{2}\operatorname{Re}\left[\ve{E}\times\ve{H}^\ast\right]=\\
&\quad\quad=\begin{cases}
\frac{1}{2\eta_{0}k_0}\left[A_\text{TE}^2k_{\text{TE},x}+A_\text{TM}^2k_{\text{TM},x}\right]\ve{\hat{x}},\\
\frac{1}{2\eta_{0}k_0}\left[A_\text{TE}^2k_{\text{TE},y}+A_\text{TM}^2k_{\text{TM},y}\right]\ve{\hat{y}},\\
\frac{A_\text{TE}A_\text{TM}}{2\eta_0k_0^2}\left[k_{\text{TE},x}k_{\text{TM},y}-k_{\text{TE},y}k_{\text{TM},x}\right]\\
\qquad\qquad\quad\cdot\cos{\left[(\ve{k}_\text{TE}-\ve{k}_\text{TM})\cdot\ve{r}-\alpha\right]}\ve{\hat{z}}.
\end{cases}
\end{split}
\end{equation}
These relations reveal a peculiar and a priori unexpected result: the superposition of these two waves leads to a non-zero transverse time-averaged Poynting vector component, i.e., $\langle S_z\rangle \neq 0$, even though $\ve{k}_\text{TE}$ and $\ve{k}_\text{TM}$ both lie within the $xy$-plane. This may seem counter-intuitive, however, it is easily explained when considering the orientation of the fields in Eqs.~\eqref{eq:EFields} and~\eqref{eq:HFields}. Indeed, due to the superposition of the two waves, the total electric and magnetic field vectors, $\ve{E}$ and $\ve{H}$, lie in a plane diagonal to the $xy$-plane, whose orientation depends on $\phi_\text{TE}$, $\phi_\text{TM}$, $A_\text{TE}$ and $A_\text{TM}$. The fact that $\langle S_z\rangle \neq 0$ directly follows from the oblique orientation of $\ve{E}$ and $\ve{H}$.

Another peculiarity of $\langle S_z\rangle $ is that it is spatially varying in a direction defined by $\ve{k}_\text{TE}-\ve{k}_\text{TM}$ and oscillates with subwavelength period $P=2\pi/|\ve{k}_\text{TE}-\ve{k}_\text{TM}|$. This implies that the spatial average of $\langle S_z\rangle$ cancels out, yielding a zero net transverse Poynting vector. However, since $\langle S_z\rangle$ remains locally non-zero, it suggests that a small particle placed within these waves may experience a non-zero net transverse force when interacting with them. We will investigate this transverse force in Secs.~\ref{sec:force} and~\ref{sec:fields} but first, we start by simplifying Eq.~\eqref{eq:Poynting}.

Upon inspection of Eq.~\eqref{eq:Poynting}, we see that $\langle S_z\rangle$ can be maximized assuming that $A_\text{TE}\neq 0$ and $A_\text{TM}\neq 0$. Its maximum value is obtained, for a given combination of $\ve{r}$ and $\alpha$,  when $|\phi_\text{TE}-\phi_\text{TM}| = (2n+1)\pi/2$, where $n\in\mathbb{Z}$. For simplicity, we thus next assume that $\phi_\text{TE} = 0$ and \mbox{$\phi_\text{TM} = \pi/2$}, as shown in Fig.~\ref{fig:Schem}, which reduces~\eqref{eq:Poynting} to
\begin{equation}
\label{eq:Poynting2}
\langle\ve{S}\rangle=\frac{1}{2\eta_{0}}
\begin{cases}
A_\text{TE}^2\ve{\hat{x}},\\
A_\text{TM}^2\ve{\hat{y}},\\
A_\text{TE}A_\text{TM}\cos{\left[k_0(x\hat{\ve{x}}-y\hat{\ve{y}})-\alpha\right]}\ve{\hat{z}}.
\end{cases}
\end{equation}
The total electric field corresponding to this particular configuration is plotted in Fig.~\ref{fig:SIM1}, where Gaussian beams are used instead of plane waves for visualization purposes. The corresponding spatially varying $\langle S_z\rangle$ is plotted in  Fig.~\ref{fig:SIM2} for $A_\text{TE}=A_\text{TM} = 1$ and $\alpha=0$. From this figure, we clearly see that $\langle S_z\rangle$ appears in the regions where the fields of the TE and TM waves overlap, which is consistent with Eq.~\eqref{eq:Poynting}.

In the next section, we demonstrate that a small particle may indeed be subjected to a transverse force when placed within these superimposed waves.

\section{Electromagnetic Forces Acting on a Sphere Illuminated by TE/TM Waves}
\label{sec:force}

\subsection{General Considerations on Electromagnetic Forces}

In the situation considered in~\eqref{eq:Poynting2}, the spatial period of $\langle S_z\rangle$ is $P=\lambda_0/\sqrt{2}$, where $\lambda_0$ is the free-space wavelength. In order to maximize the transverse force that a particle may be subjected to under such an illumination condition, we select it to be small enough so that it fits within no more than half that spatial period. Thus maximizing its interactions with the part of the beams corresponding to the transverse component of the Poynting vector. For simplicity, we consider the case of a sphere whose radius is $r_\text{s}< P/4 \approx \lambda_0/5$. 

To evaluate the forces acting on a small (subwavelength) particle, it is common practice to express them in terms of two separate distinct components~\cite{chaumetTimeaveragedTotalForce2000}. One of them being proportional to the gradient of the field intensity and the other to the absorption and scattering cross sections of the particle, $C_\text{abs}$ and $C_\text{scat}$, respectively. These two forces may be respectively expressed as~\cite{chaumetTimeaveragedTotalForce2000}
\begin{equation}
\label{eq:Fgrad}
\ve{F}_\text{grad} = \frac{1}{2}\alpha_\text{e}\ve{\nabla}|\ve{E}|^{2},
\end{equation}
and
\begin{equation}
\label{eq:FK}
\ve{F} \propto |\ve{E}|^{2}\left(C_\text{abs}+C_\text{scat}\right)\frac{\ve{k}}{k_0},
\end{equation}
where $\alpha_\text{e}$ is the electric polarizability of the particle and $\ve{k}$ is the wave vector of the illumination. While the gradient force in~\eqref{eq:Fgrad} may result in a non-zero force in the case of Gaussian beam illumination, as in Fig.~\ref{fig:SIM}, it does not yield any force when considering plane wave illumination since \mbox{$\nabla|e^{-j\ve{k}\cdot\ve{r}}|^2=0$}. On the other hand, the force given by relation~\eqref{eq:FK} leads to a non-zero force, even in the case of plane wave illumination. Indeed, for the scenario prescribed in~\eqref{eq:Poynting2}, relation~\eqref{eq:FK} results in a force oriented in the $\ve{\hat{x}}+\ve{\hat{y}}$ direction hence pushing the particle diagonally in the $xy$-plane, as illustrated by the force component $F_\text{diag}$ in Fig.~\ref{fig:Schem}. However, Eq.~\eqref{eq:FK} is clearly unable to predict the expected transverse force since it is proportional to $\ve{k}$ and not to $\ve{S}$.

Since these two simplified expressions fail to predict the existence of a transverse force, we next proceed by analyzing this problem more rigorously. For this purpose, we consider the electromagnetic conservation of momentum theorem, which reads~\cite{rothwell2008electromagnetics}
\begin{equation}
\label{eq:momentumconservation}
\ve{f} + \frac{\partial}{\partial t}\left[\ve{D}\times\ve{B}\right] = \nabla\cdot\te{T}_\text{em},
\end{equation}
where $\ve{f}$ is the volume force density and $\te{T}_\text{em}$ is the Maxwell stress tensor. The force acting on the particle may now be obtained by integrating~\eqref{eq:momentumconservation} over a fictitious volume surrounding it. Upon application of the Gauss integration law and considering that the background medium is vacuum, for which $\ve{D} = \epsilon_0 \ve{E}$ and $\ve{B} = \mu_0 \ve{H}$, Eq.~\eqref{eq:momentumconservation} becomes
\begin{equation}
\label{eq:instF}
\ve{F} = \oint_S \te{T}_\text{em} \cdot \ve{\hat{n}}~dS - \epsilon_0\mu_0\frac{\partial}{\partial t}\int_V \ve{S}~dV,
\end{equation}
where $\ve{S} = \ve{E}\times\ve{H}$ is the instantaneous Poynting vector and $\ve{\hat{n}}$ is a unit vector normal to the surface, $S$, of the integration volume, $V$. Since the last term in~\eqref{eq:instF} directly depends on the Poynting vector, it follows that the superposition of the TE and TM plane waves does generally induce a non-zero instantaneous transverse force as suggested in~\eqref{eq:Poynting2}. However, the presence of $\ve{S}$ in~\eqref{eq:instF} vanishes when considering the time-averaged force~\cite{rothwell2008electromagnetics}. Indeed, taking the average over one time period transforms~\eqref{eq:instF} into
\begin{equation}
\label{eq:avF}
\langle \ve{F} \rangle = \oint_S \langle\te{T}_\text{em}\rangle \cdot \ve{\hat{n}}~dS,
\end{equation}
where $\langle\te{T}_\text{em}\rangle$ is the time-averaged Maxwell stress tensor defined as
\begin{equation}
\label{eq:MST}
\langle\te{T}_\text{em}\rangle=\frac{1}{2}\operatorname{Re}\left[\ve{D}\ve{E}^\ast + \ve{B}\ve{H}^\ast  - \frac{1}{2} \te{I}(\ve{D}\cdot\ve{E}^\ast+\ve{B}\cdot\ve{H}^\ast)\right],
\end{equation}
with $\te{I}$ being the identity matrix. Even though~\eqref{eq:avF} represents the most general approach to investigate the average force and may thus be applied to a sphere of any size, it does not explicitly predict the existence of an average \emph{transverse} force for that system.

While it is, in some cases, possible to express~\eqref{eq:avF} directly in terms of the Poynting vector~\cite{nieto2010optical,chenOpticalPullingForce2011}, we instead consider a simplified version of~\eqref{eq:avF} that applies to electrically small particles and which provides a more explicit and insightful perspective. For that purpose, one may perform a multipolar expansion of~\eqref{eq:MST} and retain only the dipolar contributions, which are the dominant ones for an electrically small particles. This transforms~\eqref{eq:MST} into~\cite{nieto2010optical,chenOpticalPullingForce2011}
\begin{equation}
\label{eq:Fdipole}
\langle \ve{F} \rangle = \frac{1}{2}\text{Re}\left [(\nabla\ve{E}_\text{i}^\ast )\cdot\ve{p} + (\nabla\ve{H}_\text{i}^\ast )\cdot\ve{m}  -\frac{k_0^4c_0}{6\pi}(\ve{p}\times\ve{m}^\ast)\right],
\end{equation}
where $\ve{E}_\text{i}$ and $\ve{H}_\text{i}$ are the incident electric and magnetic fields, and $\ve{p}$ and $\ve{m}$ are the electric and magnetic dipole moments. They are generally defined, for an isotropic dielectric or metallic sphere, as $\ve{p} = \alpha_\text{e}\ve{E}_\text{i}$ and $\ve{m} = \alpha_\text{m}\ve{H}_\text{i}$, where ${\alpha}_\text{e}$ and ${\alpha}_\text{m}$ are the electric and magnetic susceptibilities of the particle, respectively.  

Upon inspection of~\eqref{eq:Fdipole}, we note that the two first terms reduce to~\eqref{eq:Fgrad} and~\eqref{eq:FK} for a particle with both electric and magnetic responses~\cite{chaumetTimeaveragedTotalForce2000}. As explained previously, these two terms thus predict the existence of the expected diagonal force $F_\text{diag}$ in Fig.~\ref{fig:Schem}. However, the third term on the right-hand side of~\eqref{eq:Fdipole}, which is due to the interference between the electric and magnetic dipole moments, generally yields a non-zero transverse force providing that the particle exhibits both electric and magnetic dipolar responses. 

Removing the gradient terms and substituting the dipole moments by their definition for an electrically small sphere transforms the non-vanishing term in~\eqref{eq:Fdipole} into
\begin{equation}
\label{eq:FS}
\langle \ve{F} \rangle^\text{approx} = -\frac{k_0^4c_0}{6\pi}\text{Re}\left[\alpha_\text{e}\alpha_\text{m}^\ast\right]\langle\ve{S}_\text{i}\rangle.
\end{equation}
This equation reveals an important consequence of considering the effect of both electric and magnetic responses, which is that the force may be directly related to the Poynting vector instead of the wave vector, as was the case in~\eqref{eq:FK}. Therefore, the time-averaged force acting on the particle may in general exhibit a non-zero component in a direction transverse to that of wave propagation since, in the case of the superposition of different waves, $\ve{k}$ is not necessarily parallel to $\ve{S}$.

\subsection{Comparisons Between Accurate and Approximate Force Definitions}

We shall now investigate and compare the average forces predicted by relations~\eqref{eq:avF} and~\eqref{eq:FS}. In the case of Eq.~\eqref{eq:avF}, the forces are computed directly using Mie scattering theory for a PEC sphere~\cite{paknys2016applied}. For Eq.~\eqref{eq:FS}, we use approximate expressions for the polarizabilities, which originally stem from simplified Mie coefficients~\cite{MagneticPolarizationOptical2012}. Specifically, for a sphere of radius $r_\text{s}$, relative permittivity $\epsilon_\text{r}$ and in the limiting case where $|\sqrt{\epsilon_\text{r}}k_0r_\text{s}|<1$, the polarizabilities in~\eqref{eq:FS} may be expressed as~\cite{MagneticPolarizationOptical2012}
\begin{subequations}
	\label{eq:Pola}
	\begin{align}
	\alpha_\text{e} &\approx \epsilon_04\pi r_\text{s}^3\left(\frac{\epsilon_\text{r} - 1}{\epsilon_\text{r} + 2}\right),\\
	\alpha_\text{m} &\approx \mu_04\pi r_\text{s}^3\left(\frac{r_\text{s}}{\lambda_0}\right)^2\frac{2\pi^2}{15}(\epsilon_\text{r} - 1).
	\end{align}
\end{subequations}
One may a priori think that the magnetic polarizability is negligible, which may be true for small \emph{dielectric} spheres. However, it is generally not the case, especially in the presence of important losses~\cite{MagneticPolarizationOptical2012}. Therefore, the transverse component  of the force may be negligible for dielectric particles but should still be substantial for metallic ones. As a consequence, and in order to maximize the transverse force, we next assume that the sphere is a perfect electric conductor (PEC) for which~\eqref{eq:Pola} may be expressed as $\alpha_\text{e} \approx \epsilon_0 4\pi r_\text{s}^3$ and $\alpha_\text{m} \approx -\mu_0 4\pi r_\text{s}^3$ (quasi-static approximation)~\cite{jackson_classical_1999}.

We now investigate the forces acting on this PEC sphere using two different methods: an accurate one, based on rigorous Mie scattering functions along with the Maxwell stress tensor in Eq.~\eqref{eq:MST}, and an approximate one based on Eq.~\eqref{eq:FS}. To clearly highlight the differences between these methods, we consider a radius-to-wavelength ratio ($r_\text{s}/\lambda_0$) ranging from 0 to 4, which purposefully exceeds the validity range of Eq.~\eqref{eq:FS}. 

\begin{figure}[h!]
	\centering
	\includegraphics[width=1\linewidth]{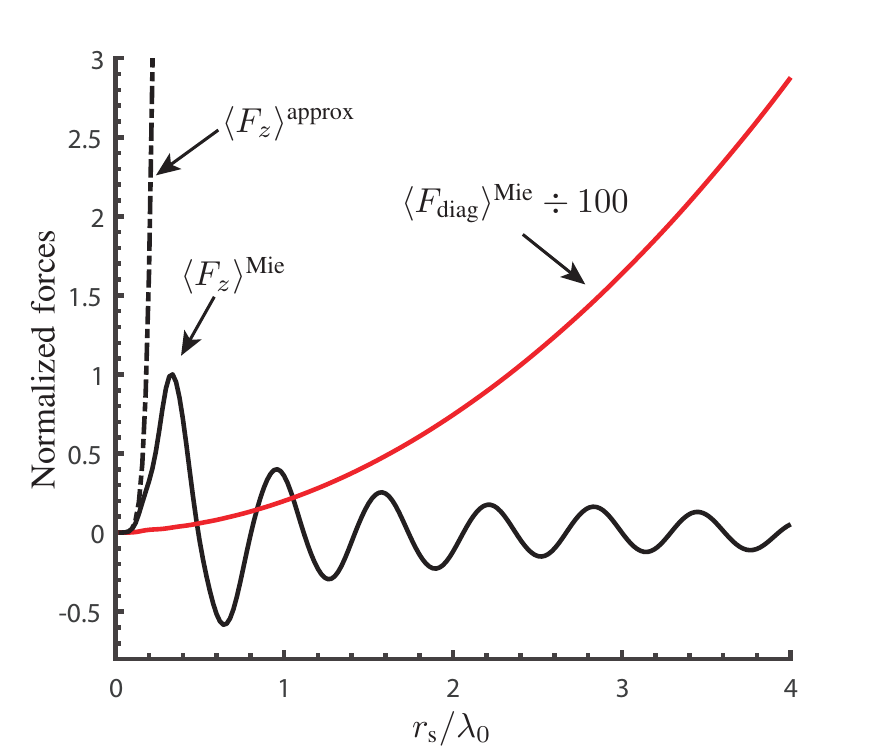}
	\caption{Forces acting on a PEC sphere with varying radius to wavelength ratio and illuminated by a TE and a TM plane waves, as specified in~\eqref{eq:Poynting2}. These forces are normalized with respect to the maximum of $\langle F_{z}\rangle^\text{Mie}$. $\langle F_{z}\rangle^\text{approx}$ and $\langle F_{z}\rangle^\text{Mie}$ are the transverse forces predicted by Eq.~\eqref{eq:FS} and Mie theory, respectively. $\langle F_\text{diag}\rangle^\text{Mie}$ is the in-plane diagonal force obtained with Mie theory.} 
	\label{fig:forces}
\end{figure}

Assuming the same illumination condition as in~\eqref{eq:Poynting2}, we plot our results in Fig.~\ref{fig:forces}, where $\langle F_z\rangle^\text{approx}$ (dashed black line) is the average transverse force given by~\eqref{eq:FS} with the polarizabilities of a PEC sphere, $\langle F_z\rangle^\text{Mie}$ (solid black line) and  $\langle F_\text{diag}\rangle^\text{Mie}$ (solid red line) are the average transverse and in-plane diagonal forces given by~\eqref{eq:avF} using Mie theory, respectively. Here, $\langle F_\text{diag}\rangle^\text{Mie}$ thus refers to the force oriented in the $\ve{\hat{x}}+\ve{\hat{y}}$ direction and is the equivalent of the force predicted by~\eqref{eq:FK}.

As expected, $\langle F_\text{diag}\rangle^\text{Mie}$ exhibits a strongly nonlinear increase, proportional to $r_\text{s}^6$, and clearly dominates for spheres larger than the wavelength. More interesting is the transverse force $\langle F_{z}\rangle^\text{Mie}$ that exhibits an oscillating behavior, which damps out as function of $r_\text{s}/\lambda_0$. Regarding the approximate transverse force, $\langle F_z\rangle^\text{approx}$, we can see that it provides a good approximation of $\langle F_z\rangle^\text{Mie}$ in the limit up to $r_\text{s}/\lambda_0< 0.1$, which is expected considering the assumptions made to derive~\eqref{eq:FS}. Beyond that limit, it diverges and thus fails to predict both the oscillating and the dampening behaviors that $\langle F_z\rangle^\text{Mie}$ exhibits. In order to better understand the physics behind these behaviors, we provide in the next section a multipolar analysis of the origin of $\langle F_z\rangle^\text{Mie}$.

\section{Multipolar Origin of the Transverse Force}
\label{sec:fields}

\begin{figure}[h!]
	\centering
	\includegraphics[width=1\linewidth]{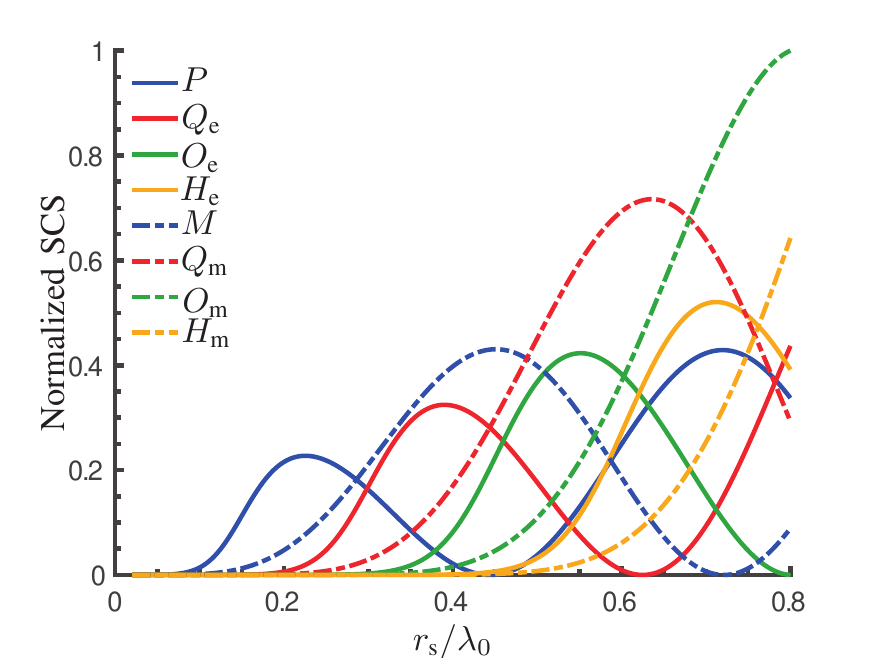}
	\caption{Normalized scattering cross sections corresponding to electric and magnetic dipolar ($P$ and $M$), quadrupolar ($Q_\text{e/m}$), octopolar ($O_\text{e/m}$) and hexadecapolar ($H_\text{e/m}$) components, respectively.}
	\label{fig:DQOH}
\end{figure}

We shall now concentrate our attention on the transverse force  and investigate its origin using a multipolar based analysis. Even though Eq.~\eqref{eq:FS} already provides a hint on the origin of the transverse force, namely that the latter emerges from the combined effects of electric and magnetic dipolar responses, it is not sufficient to fully understand the behavior of $\langle F_z\rangle^\text{Mie}$. This is because the scattering from the sphere requires a plethora of multipolar modes to be properly assessed even for relatively small values of $r_\text{s}/\lambda_0$, which, as explained above, is missing in~\eqref{eq:FS}. To demonstrate this, we have plotted, in Fig.~\ref{fig:DQOH}, the scattering cross section corresponding to the first 4 electric (solid lines) and magnetic (dashed lines) multipoles. Note that, for convenience, we have decided to restrict our attention to the range $0 \leq r_\text{s}/\lambda_0 \leq 0.8$ to avoid overcrowding the plot. 

Figure~\ref{fig:DQOH} clearly shows the complexity of the problem and confirms that many multipolar modes should be taken into account to properly approximate the scattering behavior of the sphere. However, our goal is not to provide an accurate approximation of the transverse force but rather to explain its physical origin. For that purpose, and in order to avoid lengthy considerations, we next concentrate our efforts on the dipolar and quadrupolar contributions. Accordingly, we now further analyze the curves in Fig.~\ref{fig:DQOH} so as to reveal which of the dipolar and quadrupolar components are induced by the TE and TM plane waves, respectively. Since the sphere is isotropic, the dipolar components can be directly associated to the orientation of the electric and magnetic fields of the waves. The quadrupolar components may be found as $Q_{\text{e}, ij} \propto \partial_i E_j + \partial_j E_i$ and $Q_{\text{m}, ij} \propto \partial_i H_j + \partial_j H_i$ and noting that $Q_{\text{e/m}, ij} = Q_{\text{e/m}, ji}$, where $\partial_{i/j}$ represent the partial derivatives with respect to $i, j = \{x, y,z\}$~\cite{aluGuidedPropagationQuadrupolar2009,bernalarangoUnderpinningHybridizationIntuition2014a,mobiniTheoryOpticalForces2018}. Due to the polarization state and direction of propagation of the two waves, we can conclude that the TE wave induces the components $p_z, m_y, Q_{\text{e},xz}$ and $Q_{\text{m},xy}$, while the TM wave induces the components $p_x, m_z, Q_{\text{e},xy}$ and $Q_{\text{m},yz}$. One may refer to Fig.~\ref{fig:Scat} to see the radiation patterns corresponding to these multipoles. 

Since the sphere is made out of PEC, it does not absorb the energy of the incident waves. Therefore, the forces acting on the sphere are strictly due to the fields that it scatters~\cite{nieto2010optical,chenOpticalPullingForce2011,mobiniTheoryOpticalForces2018}. This implies that these forces cannot result from the contribution of individual multipolar components, due to their symmetric radiation pattern, but rather from the superposition of at least two multipolar contributions, as explained in~\cite{nieto2010optical,chenOpticalPullingForce2011,mobiniTheoryOpticalForces2018}. Hence, it follows that the total force acting on the sphere may be written as the sum of these multipolar contributions as~\cite{chenOpticalPullingForce2011,salandrinoGeneralizedMieTheory2012,mobiniTheoryOpticalForces2018}
\begin{equation}
\label{eq:forceDec}
\langle \ve{F}\rangle^\text{Mie} = \langle \ve{F}\rangle^\text{PM} + \langle \ve{F}\rangle^{\text{PQ}_\text{e}} + \langle \ve{F}\rangle^{\text{MQ}_\text{m}} + \langle \ve{F}\rangle^{\text{Q}_\text{e}\text{Q}_\text{m}} + \dots
\end{equation}
where $\langle \ve{F}\rangle^\text{PM}$ represents the time-averaged force due to the superposition of the scattering from the sphere electric and magnetic dipolar responses, $\langle \ve{F}\rangle^{\text{PQ}_\text{e}}$ represents the force corresponding to the superposition of electric dipolar and quadrupolar responses, and so on.

\begin{figure}[h!]
	\centering
	\includegraphics[width=1\linewidth]{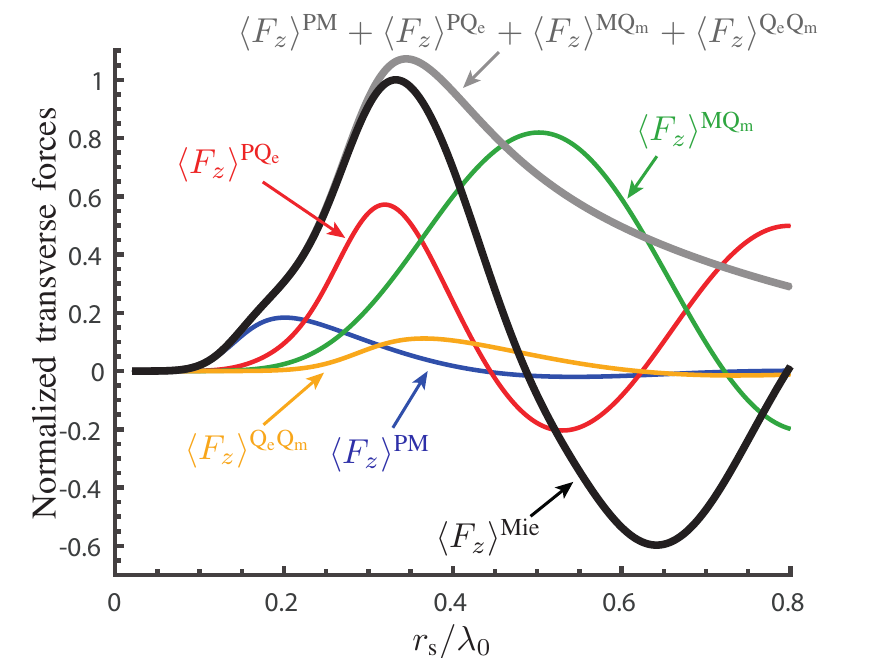}
	\caption{Decomposition of $\langle F_{z}\rangle^\text{Mie}$ into dipolar and quadrupolar contributions. These forces are normalized with respect to the maximum of $\langle F_{z}\rangle^\text{Mie}$.}
	\label{fig:F_p_m_q}
\end{figure}

Since we know the total electromagnetic fields scattered by the sphere from Mie theory, we next compute each component in~\eqref{eq:forceDec} by expressing the dipolar and quadrupolar responses directly in terms of the corresponding Mie coefficients~\cite{muhligMultipoleAnalysisMetaatoms2011}. We then isolate the $z$-oriented component of the resulting force and, for each term in~\eqref{eq:forceDec}, plot their corresponding behavior in Fig.~\ref{fig:F_p_m_q}. 
\setcounter{figure}{5}
In this figure, the black curve corresponds to $\langle F_{z}\rangle^\text{Mie}$, the same as in Fig.~\ref{fig:forces}, and the gray curve represents the approximation of $\langle F_{z}\rangle^\text{Mie}$ given by~\eqref{eq:forceDec} in terms of dipolar and quadrupolar modes. As can be seen in Fig.~\ref{fig:forces}, Eq.~\eqref{eq:forceDec} is in a good agreement with $\langle F_{z}\rangle^\text{Mie}$ up to $r_\text{s}/\lambda_0 \approx 0.2$. For larger radius-to-wavelength ratios, it starts to diverge, which is to be expected since, referring to Fig.~\ref{fig:DQOH}, it is where the higher-order modes (such as $O_\text{e}$) start to play a role. We also note that the behavior of $\langle F_z\rangle^\text{PM}$ in Fig.~\ref{fig:F_p_m_q} strongly differs from that given by~\eqref{eq:FS} and plotted as $\langle F_{z}\rangle^\text{approx}$ in Fig.~\ref{fig:forces}. Indeed, even though they both represent the force due to the combined effects of electric and magnetic dipolar responses, Eq.~\eqref{eq:FS} was specifically derived for the case of very small values of $r_\text{s}/\lambda_0$, while $\langle F_z\rangle^\text{PM}$ in Fig.~\ref{fig:F_p_m_q} is obtained directly from Mie theory and thus applies to any value of $r_\text{s}/\lambda_0$.

\begin{figure*}[t]
	\centering
	\includegraphics[width=1\linewidth]{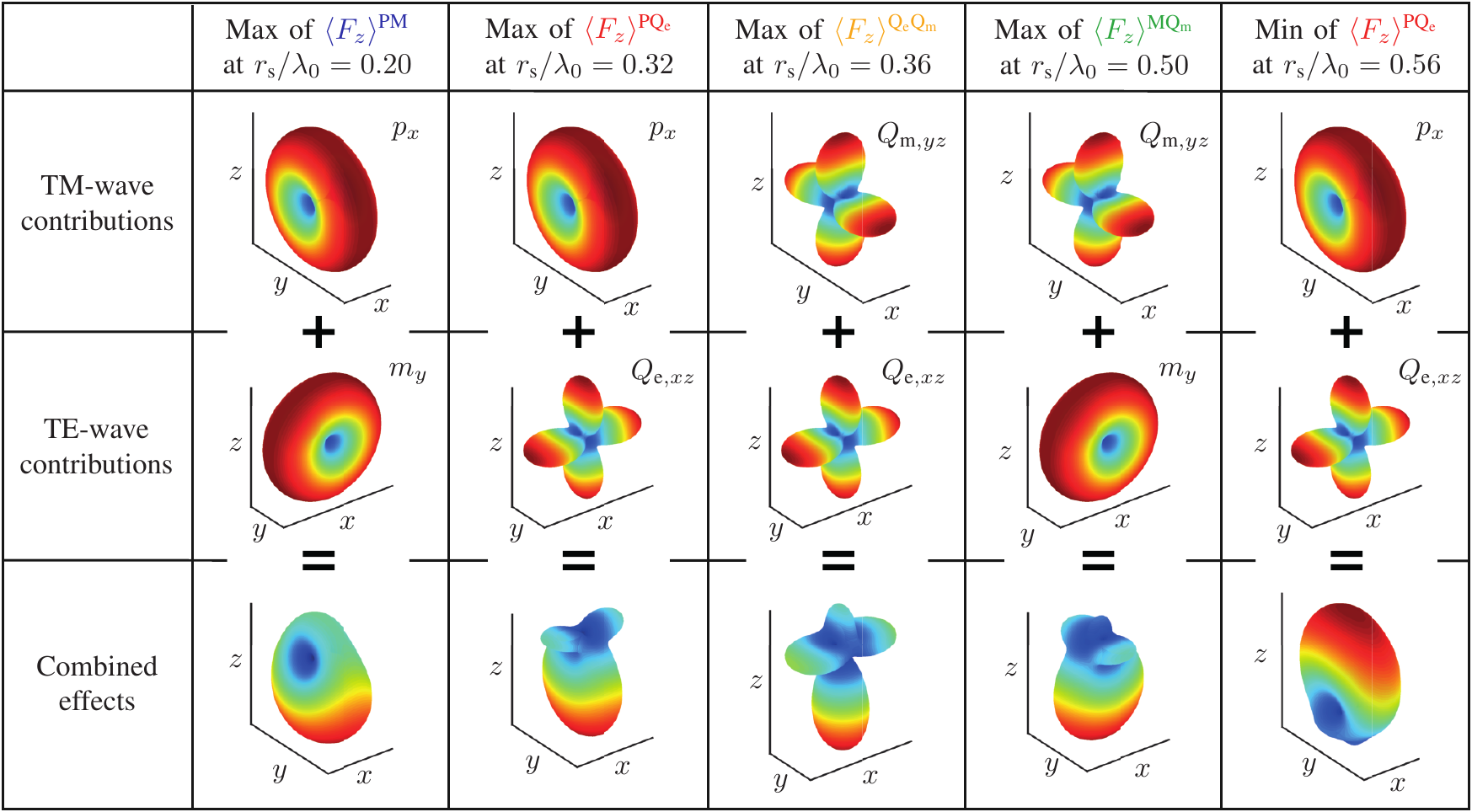}
	\caption{Radiation patterns of the dipolar and quadrupolar components that contribute the transverse force.}
	\label{fig:Scat}
\end{figure*}

A close inspection of the 4 components of the force, given in~\eqref{eq:forceDec} and that are plotted in Fig.~\ref{fig:F_p_m_q}, reveals that they all exhibit maxima, minima and nodes depending on the value of $r_\text{s}/\lambda_0$. To better understand the origin of these oscillations, we next investigate the field interactions of the corresponding multipolar modes for these different cases. Since we are here only interested in the transverse force, we can ignore the multipolar components that do not contribute to it. Indeed, the force $\langle \ve{F}\rangle^\text{PM}$ may be split into
\begin{equation}
\langle \ve{F}\rangle^\text{PM} = \langle \ve{F}\rangle^{p_x m_y} + \langle \ve{F}\rangle^{p_x m_z} + \langle \ve{F}\rangle^{p_z m_y}+ \langle \ve{F}\rangle^{p_z m_z}.
\end{equation}
Out of these contributions, only the superposition of $p_x$ and $m_y$ in $\langle \ve{F}\rangle^{p_x m_y}$ contributes to the transverse force, while the other ones result in a zero net transverse force because of their symmetric radiation pattern through the $xy$-plane (mirror symmetry). Note that we assume here that there exists a direct relationship between the electromagnetic force and the intensity of the fields. This is indeed the case since, in the far-field, we have that $\ve{E}\times\ve{H} \parallel \ve{\hat{n}}$ implying that~\eqref{eq:avF} reduces to $\langle \ve{F} \rangle = -\frac{\epsilon_0}{2} \oint_S |\ve{E}|^2 \cdot \ve{\hat{n}}~dS$. By applying the same consideration of scattering symmetry, the other force components in~\eqref{eq:forceDec} are reduced such that the transverse component of~\eqref{eq:forceDec} becomes
\begin{equation}
\label{eq:Fzcomp}
\begin{split}
\langle {F}_z\rangle^\text{Mie} \approx & \langle {F}_z\rangle^{p_x m_y} + \langle {F}_z\rangle^{p_x Q_{\text{e},xz}} \\
&\qquad+ \langle {F}_z\rangle^{m_y Q_{\text{m},yz}} + \langle {F}_z\rangle^{{Q}_{\text{e},xz}{Q}_{\text{m},yz}}.
\end{split}
\end{equation}
In order to visualize how the fields from these modes combine together to yield a non-zero net transverse force, we plot in Fig.~\ref{fig:Scat} the radiation pattern corresponding to the first maximum of $\langle {F}_z\rangle^{p_x m_y}$, $\langle {F}_z\rangle^{p_x Q_{\text{e},xz}}$, $\langle {F}_z\rangle^{m_y Q_{\text{m},yz}}$ and $\langle {F}_z\rangle^{{Q}_{\text{e},xz}{Q}_{\text{m},yz}}$ as well as the first minimum of  $\langle {F}_z\rangle^{p_x Q_{\text{e},xz}}$.

These plots clearly demonstrate that the transverse force is due to the combined effects of the TE and TM waves that results in asymmetric radiation patterns. This is because the fields scattered by these multipolar components interfere constructively/destructively in the $\pm z$-directions depending on the relative phase shift between the corresponding modes in a way that is reminiscent of the Kerker effects~\cite{Kerker2013,alaeeGeneralizedKerkerCondition2015}. Hence, a positive transverse force is due to the constructive interference of the modes in the $-z$-direction and their destructive interference in the $+z$-direction, resulting in more power being scattered downward, thus pushing the particle upward by conservation of momentum. Obviously, the opposite occurs for the minima of the force, as illustrated in the last column in Fig.~\ref{fig:Scat}, where the interference of the modes leads to significant upward scattering thus resulting in a negative force.

\begin{figure}[h!]
	\centering
	\includegraphics[width=1\linewidth]{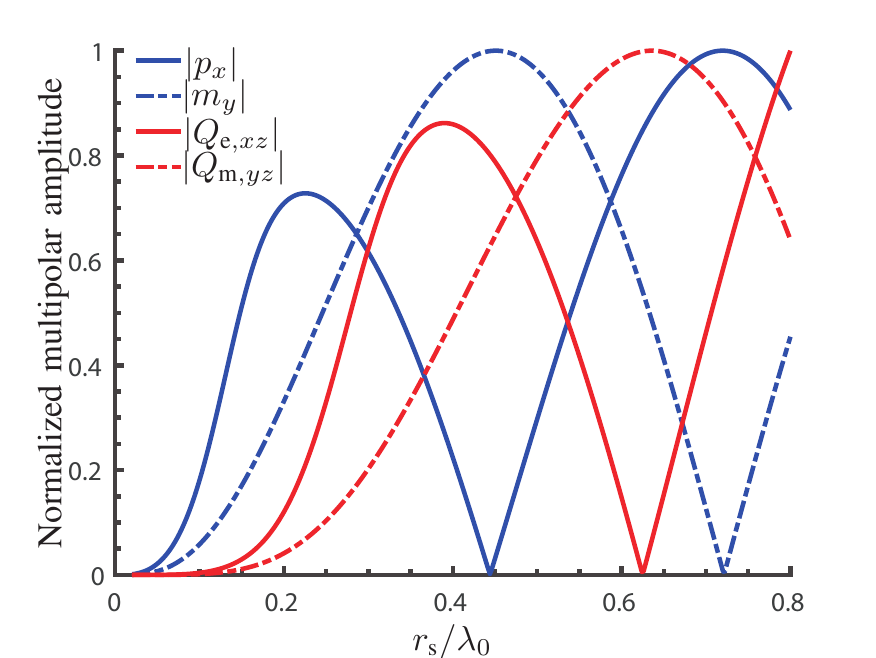}
	\caption{Amplitude of the relevant dipolar and quadrupolar components each individually normalized with respect to their respective maximum. The three zeros are at $r_\text{s}/\lambda_0 = \{0.44, 0.62, 0.72\}$, respectively.}
	\label{fig:PQ}
\end{figure}

The oscillating behavior that $\langle F_{z}\rangle^\text{Mie}$ exhibits as function of $r_\text{s}/\lambda_0$ directly stems from these interference effects and is explained by the fact that the respective amplitude and phase of each mode vary differently as the value of $r_\text{s}/\lambda_0$ changes, hence leading to the observed maxima and minima. Accordingly, one may a priori think that the nodes of the components of the transverse force in~\eqref{eq:forceDec} would stem from a balance of the modes, in terms of their relative amplitude and phase, resulting in symmetric radiation patterns. However, the actual origin of these nodes is the fact that, for specific values of $r_\text{s}/\lambda_0$, the amplitude of one of the two involved mode components is zero, as evidenced by the curves in Fig.~\ref{fig:PQ} that correspond to the amplitude of $p_x$, $m_y$, $Q_{\text{e},xz}$ and $Q_{\text{m},yz}$.
This is indeed verified by comparing the values of the zeros in Fig.~\ref{fig:PQ} with the values of the nodes of the force components in Fig.~\ref{fig:F_p_m_q}. For instance, we can see that the first node of $\langle {F}_z\rangle^{p_x m_y}$ and  $\langle {F}_z\rangle^{p_x Q_{\text{e},xz}}$ is at $r_\text{s}/\lambda_0 = 0.44$, i.e., when $|p_x| = 0$. Similarly, the first node of $\langle {F}_z\rangle^{{Q}_{\text{e},xz}{Q}_{\text{m},yz}}$ is at $r_\text{s}/\lambda_0 = 0.62$, i.e., when $|{Q}_{\text{e},xz}| = 0$ and that of $\langle {F}_z\rangle^{m_y Q_{\text{m},yz}}$ is at $r_\text{s}/\lambda_0 = 0.72$, i.e., when $|m_y| = 0$, and so on.

\begin{figure}[h!]
	\centering
	\includegraphics[width=1\linewidth]{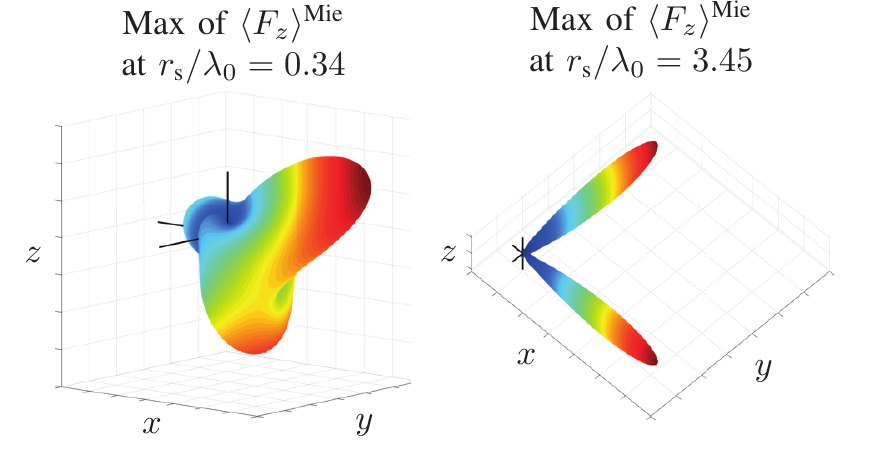}
	\caption{Total radiation patterns of the sphere when illuminated by the TE and TM plane waves. The intersection of the black lines indicates the position of the sphere.}
	\label{fig:rad}
\end{figure}

Now that we have explained the reasons for the oscillating behavior of the transverse force, we investigate the decaying trend that $\langle F_z\rangle^\text{Mie}$ exhibits as the value of $r_\text{s}/\lambda_0$ increases, which is clearly visible in Fig.~\ref{fig:forces}. This turns out to be a consequence of the typical asymmetric scattering behavior that large spheres exhibit. This asymmetry, which may be characterized by an ``asymmetry parameter'' defined as the average cosine of the scattering angle~\cite{vandehulstLightScatteringSmall,bohrenAbsorptionScatteringLight1983}, results from the constructive interference of the fields scattered by the excited multipoles, which tends to produce a strong forward scattering, and becomes more and more dominant as the value of $r_\text{s}/\lambda_0$ increases. This situation is illustrated in Fig.~\ref{fig:rad}, where the total (TE + TM waves) radiation patterns are plotted when $r_\text{s}/\lambda_0 = \{0.34, 3.45\}$ corresponding to two maxima of $\langle F_z\rangle^\text{Mie}$ (refer to Fig.~\ref{fig:forces}).

We can see from these two figures that the radiation pattern at $r_\text{s}/\lambda_0 = 0.34$, being dominated by dipolar and quadrupolar modes, is more angularly symmetric than the other one, which exhibits two important scattering lobes oriented in the direction of propagation of the TE and TM waves, respectively. It follows that, as the value of $r_\text{s}/\lambda_0$ increases, more power is scattered in the direction of the illumination and thus less power is available to produce a transverse force, which results in the decay of $\langle F_z\rangle^\text{Mie}$ that is visible in Fig.~\ref{fig:forces}.

\section{Conclusion}

We have investigated the electromagnetic forces acting on a metallic sphere illumined by the superposition of a TE and TM plane waves propagating in the $+x$- and $+y$-directions, respectively. In particular, we have concentrated our attention on the transverse component of this force as its origin remains counter-intuitive and difficult to explain theoretically. We have first seen that a tentative expression of the force based on a small-particle approximation, while able to predict the existence of a transverse force, fails to capture its dynamic as the size of the particle increases compared to the wavelength. To overcome this limitation, we have proposed a multipolar-based analysis aimed at providing an intelligible explanation of that phenomenon. This analysis has revealed that the transverse force stems from the constructive and destructive interference of the fields scattered by the multipoles induced by the combined illuminations. It follows that the observed oscillations of the transverse force, i.e., its maxima, minima, nodes and decay as function of the radius-to-wavelength ratio, are all a consequence of these interference effects.

\section*{Acknowledgments}

We gratefully acknowledge funding from the European Research Council (ERC-2015-AdG-695206 Nanofactory).

\bibliographystyle{myIEEEtran}
\bibliography{NewLib}

\begin{thebibliography}{10}
\providecommand{\url}[1]{#1}
\csname url@samestyle\endcsname
\providecommand{\newblock}{\relax}
\providecommand{\bibinfo}[2]{#2}
\providecommand{\BIBentrySTDinterwordspacing}{\spaceskip=0pt\relax}
\providecommand{\BIBentryALTinterwordstretchfactor}{4}
\providecommand{\BIBentryALTinterwordspacing}{\spaceskip=\fontdimen2\font plus
\BIBentryALTinterwordstretchfactor\fontdimen3\font minus
  \fontdimen4\font\relax}
\providecommand{\BIBforeignlanguage}[2]{{%
\expandafter\ifx\csname l@#1\endcsname\relax
\typeout{** WARNING: IEEEtran.bst: No hyphenation pattern has been}%
\typeout{** loaded for the language `#1'. Using the pattern for}%
\typeout{** the default language instead.}%
\else
\language=\csname l@#1\endcsname
\fi
#2}}
\providecommand{\BIBdecl}{\relax}
\BIBdecl

\bibitem{jonasLightWorkUse2008}
A.~Jon{\'a}{\v s} and P.~Zem{\'a}nek, ``Light at work: {{The}} use of optical
  forces for particle manipulation, sorting, and analysis,''
  \emph{Electrophoresis}, vol.~29, no.~24, pp. 4813--4851, Dec. 2008.

\bibitem{zemanekPerspectiveLightinducedTransport2019}
P.~Zem{\'a}nek, G.~Volpe, A.~Jon{\'a}{\v s}, and O.~Brzobohat{\'y},
  ``Perspective on light-induced transport of particles: From optical forces to
  phoretic motion,'' \emph{Advances in Optics and Photonics}, vol.~11, no.~3,
  p. 577, Sep. 2019.

\bibitem{ashkinAccelerationTrappingParticles1970}
A.~Ashkin, ``Acceleration and {{Trapping}} of {{Particles}} by {{Radiation
  Pressure}},'' \emph{Physical Review Letters}, vol.~24, no.~4, pp. 156--159,
  Jan. 1970.

\bibitem{ashkin1986observation}
A.~Ashkin, J.~M. Dziedzic, J.~Bjorkholm, and S.~Chu, ``Observation of a
  single-beam gradient force optical trap for dielectric particles,''
  \emph{Optics letters}, vol.~11, no.~5, pp. 288--290, 1986.

\bibitem{neumanOpticalTrapping2004}
K.~C. Neuman and S.~M. Block, ``Optical trapping,'' \emph{Review of Scientific
  Instruments}, vol.~75, no.~9, pp. 2787--2809, Sep. 2004.

\bibitem{wangMicrofluidicSortingMammalian2005}
M.~M. Wang, E.~Tu, D.~E. Raymond, J.~M. Yang, H.~Zhang, N.~Hagen, B.~Dees,
  E.~M. Mercer, A.~H. Forster, I.~Kariv, P.~J. Marchand, and W.~F. Butler,
  ``Microfluidic sorting of mammalian cells by optical force switching,''
  \emph{Nature Biotechnology}, vol.~23, no.~1, pp. 83--87, Jan. 2005.

\bibitem{yangOpticalManipulationNanoparticles2009}
A.~H.~J. Yang, S.~D. Moore, B.~S. Schmidt, M.~Klug, M.~Lipson, and D.~Erickson,
  ``Optical manipulation of nanoparticles and biomolecules in sub-wavelength
  slot waveguides,'' \emph{Nature}, vol. 457, no. 7225, pp. 71--75, Jan. 2009.

\bibitem{juan2011plasmon}
M.~L. Juan, M.~Righini, and R.~Quidant, ``Plasmon nano-optical tweezers,''
  \emph{Nature photonics}, vol.~5, no.~6, p. 349, 2011.

\bibitem{dholakia2011shaping}
K.~Dholakia and T.~{\v{C}}i{\v{z}}m{\'a}r, ``Shaping the future of
  manipulation,'' \emph{Nature Photonics}, vol.~5, no.~6, p. 335, 2011.

\bibitem{pang2012optical}
Y.~Pang and R.~Gordon, ``Optical trapping of a single protein,'' \emph{Nano
  letters}, vol.~12, no.~1, pp. 402--406, 2012.

\bibitem{Johnson2011}
L.~Johnson, M.~Whorton, A.~Heaton, R.~Pinson, G.~Laue, and C.~Adams,
  ``Nanosail-d: A solar sail demonstration mission,'' \emph{Acta Astronaut.},
  vol.~68, no.~5, pp. 571--575, 2011.

\bibitem{achouriSolarMetaSailsAgile2019}
K.~Achouri, O.~V. Cespedes, and C.~Caloz, ``Solar ``{{Meta}}-{{Sails}}'' for
  {{Agile Optical Force Control}},'' \emph{IEEE Transactions on Antennas and
  Propagation}, vol.~67, no.~11, pp. 6924--6934, Nov. 2019.

\bibitem{swartzlanderLightSailingGreat2020}
G.~Swartzlander, L.~Johnson, and B.~Betts, ``Light {{Sailing}} into the {{Great
  Beyond}},'' \emph{Optics and Photonics News}, vol.~31, no.~2, p.~30, Feb.
  2020.

\bibitem{radescuExactCalculationAngular2002}
E.~E. Radescu and G.~Vaman, ``Exact calculation of the angular momentum loss,
  recoil force, and radiation intensity for an arbitrary source in terms of
  electric, magnetic, and toroid multipoles,'' \emph{Physical Review E},
  vol.~65, no.~4, p. 046609, Apr. 2002.

\bibitem{rothwell2008electromagnetics}
E.~J. Rothwell and M.~J. Cloud, \emph{Electromagnetics}.\hskip 1em plus 0.5em
  minus 0.4em\relax CRC press, 2008.

\bibitem{salandrinoGeneralizedMieTheory2012}
A.~Salandrino, S.~Fardad, and D.~N. Christodoulides, ``Generalized {{Mie}}
  theory of optical forces,'' \emph{Journal of the Optical Society of America
  B}, vol.~29, no.~4, p. 855, Apr. 2012.

\bibitem{Novotny2012}
L.~Novotny and B.~Hecht, \emph{Principles of nano-optics}.\hskip 1em plus 0.5em
  minus 0.4em\relax Cambridge university press, 2012.

\bibitem{chenOpticalPullingForce2011}
J.~Chen, J.~Ng, Z.~Lin, and C.~T. Chan, ``Optical pulling force,'' \emph{Nature
  Photonics}, vol.~5, no.~9, pp. 531--534, Sep. 2011.

\bibitem{bliokhExtraordinaryMomentumSpin2014}
K.~Y. Bliokh, A.~Y. Bekshaev, and F.~Nori, ``Extraordinary momentum and spin in
  evanescent waves,'' \emph{Nature Communications}, vol.~5, no.~1, p. 3300, May
  2014.

\bibitem{bekshaevTransverseSpinMomentum2015}
A.~Y. Bekshaev, K.~Y. Bliokh, and F.~Nori, ``Transverse {{Spin}} and
  {{Momentum}} in {{Two}}-{{Wave Interference}},'' \emph{Physical Review X},
  vol.~5, no.~1, p. 011039, Mar. 2015.

\bibitem{bliokhTransverseLongitudinalAngular2015}
K.~Y. Bliokh and F.~Nori, ``Transverse and longitudinal angular momenta of
  light,'' \emph{Physics Reports}, vol. 592, pp. 1--38, Aug. 2015.

\bibitem{antognozziDirectMeasurementsExtraordinary2016}
M.~Antognozzi, C.~R. Bermingham, R.~L. Harniman, S.~Simpson, J.~Senior,
  R.~Hayward, H.~Hoerber, M.~R. Dennis, A.~Y. Bekshaev, K.~Y. Bliokh, and
  F.~Nori, ``Direct measurements of the extraordinary optical momentum and
  transverse spin-dependent force using a nano-cantilever,'' \emph{Nature
  Physics}, vol.~12, no.~8, pp. 731--735, Aug. 2016.

\bibitem{mobiniTheoryOpticalForces2018}
E.~Mobini, A.~Rahimzadegan, C.~Rockstuhl, and R.~Alaee, ``Theory of optical
  forces on small particles by multiple plane waves,'' \emph{Journal of Applied
  Physics}, vol. 124, no.~17, p. 173102, Nov. 2018.

\bibitem{gurvitzNumericalCalculationCartesian2018}
E.~Gurvitz and A.~S. Shalin, ``Numerical calculation and {{Cartesian}}
  multipole decomposition of optical pulling force acting on {{Si}} nanocube in
  visible region,'' \emph{Journal of Physics: Conference Series}, vol. 1092, p.
  012048, Sep. 2018.

\bibitem{nieto2010optical}
M.~Nieto-Vesperinas, J.~S{\'a}enz, R.~G{\'o}mez-Medina, and L.~Chantada,
  ``Optical forces on small magnetodielectric particles,'' \emph{Optics
  express}, vol.~18, no.~11, pp. 11\,428--11\,443, 2010.

\bibitem{evlyukhinCollectiveResonancesMetal2012}
A.~B. Evlyukhin, C.~Reinhardt, U.~Zywietz, and B.~N. Chichkov, ``Collective
  resonances in metal nanoparticle arrays with dipole-quadrupole
  interactions,'' \emph{Physical Review B}, vol.~85, no.~24, p. 245411, Jun.
  2012.

\bibitem{chaumetTimeaveragedTotalForce2000}
P.~C. Chaumet and M.~{Nieto-Vesperinas}, ``Time-averaged total force on a
  dipolar sphere in an electromagnetic field,'' \emph{Optics Letters}, vol.~25,
  no.~15, p. 1065, Aug. 2000.

\bibitem{paknys2016applied}
R.~Paknys, \emph{Applied frequency-domain electromagnetics}.\hskip 1em plus
  0.5em minus 0.4em\relax John Wiley \& Sons, 2016.

\bibitem{MagneticPolarizationOptical2012}
A.~{Asenjo-Garcia}, A.~Manjavacas, V.~Myroshnychenko, and F.~J. {Garc{\'i}a de
  Abajo}, ``Magnetic polarization in the optical absorption of metallic
  nanoparticles,'' \emph{Optics Express}, vol.~20, no.~27, p. 28142, Dec. 2012.

\bibitem{jackson_classical_1999}
J.~D. Jackson, \emph{Classical electrodynamics}, 3rd~ed.\hskip 1em plus 0.5em
  minus 0.4em\relax New York, {NY}: Wiley, 1999.

\bibitem{aluGuidedPropagationQuadrupolar2009}
A.~Al{\`u} and N.~Engheta, ``Guided propagation along quadrupolar chains of
  plasmonic nanoparticles,'' \emph{Physical Review B}, vol.~79, no.~23, p.
  235412, Jun. 2009.

\bibitem{bernalarangoUnderpinningHybridizationIntuition2014a}
F.~Bernal~Arango, T.~Coenen, and A.~F. Koenderink, ``Underpinning
  {{Hybridization Intuition}} for {{Complex Nanoantennas}} by {{Magnetoelectric
  Quadrupolar Polarizability Retrieval}},'' \emph{ACS Photonics}, vol.~1,
  no.~5, pp. 444--453, May 2014.

\bibitem{muhligMultipoleAnalysisMetaatoms2011}
S.~M{\"u}hlig, C.~Menzel, C.~Rockstuhl, and F.~Lederer, ``Multipole analysis of
  meta-atoms,'' \emph{Metamaterials}, vol.~5, no. 2-3, pp. 64--73, Jun. 2011.

\bibitem{Kerker2013}
M.~Kerker, \emph{The scattering of light and other electromagnetic radiation:
  physical chemistry: a series of monographs}.\hskip 1em plus 0.5em minus
  0.4em\relax Academic press, 2013, vol.~16.

\bibitem{alaeeGeneralizedKerkerCondition2015}
R.~Alaee, R.~Filter, D.~Lehr, F.~Lederer, and C.~Rockstuhl, ``A generalized
  {{Kerker}} condition for highly directive nanoantennas,'' \emph{Optics
  Letters}, vol.~40, no.~11, p. 2645, Jun. 2015.

\bibitem{vandehulstLightScatteringSmall}
H.~C. Hulst and H.~C. van~de Hulst, \emph{Light scattering by small
  particles}.\hskip 1em plus 0.5em minus 0.4em\relax Courier Corporation, 1981.

\bibitem{bohrenAbsorptionScatteringLight1983}
C.~F. Bohren and D.~R. Huffman, \emph{Absorption and Scattering of Light by
  Small Particles}.\hskip 1em plus 0.5em minus 0.4em\relax {New York}: {Wiley},
  1983.

\end{thebibliography}

\end{document}